\documentclass[preprint,3p,12pt,pdf]{elsarticle}

\usepackage{lineno,hyperref}
\modulolinenumbers[5]

\usepackage{hyperref}

\usepackage{epstopdf}

\usepackage{tikz}
\usetikzlibrary{arrows}

\usepackage{pst-node}
\usepackage{tikz-cd}

\usepackage[shellescape]{gmp}

\usepackage{mathrsfs}
\usepackage{amssymb}
\usepackage{amsfonts}
\usepackage{latexsym}
\usepackage{amsmath}
\usepackage{xcolor}
\usepackage{wrapfig}
\usepackage{floatflt}

\usepackage{mathtools}
\usepackage{extarrows}

\usepackage{graphicx}

\usepackage{subcaption}

\def\lb{\label}

\newcommand{\er}[1]{\textrm{(\ref{#1})}}

\newtheorem{theorem}{\bf Theorem}[section]

\def\d{\delta}         
\def\D{\Delta}

 \def\cM{{\mathcal M}}       
 \def\cN{{\mathcal N}}

\def\o{\omega}

\def\Z{{\mathbb Z}}    \def\R{{\mathbb R}}   \def\C{{\mathbb C}}

\def\lt{\biggl}                  \def\rt{\biggr}
               \def\wt{\widetilde}

                 \let\le\leqslant

\def\iy{\infty}
\def\sm{\setminus}               \def\es{\emptyset}
\def\ss{\subset}                 \def\ts{\times}
                
                 \def\ev{\equiv}
        
\def\el2{\ell^{\,2}}             \def\1{1\!\!1}

\def\diag{\mathop{\mathrm{diag}}\nolimits}

\def\Im{\mathop{\mathrm{Im}}\nolimits}

\let\le\leqslant

\newcommand{\ca}{\begin{cases}}
\newcommand{\ac}{\end{cases}}
\newcommand{\ma}{\begin{pmatrix}}
\newcommand{\am}{\end{pmatrix}}
\def\eq{\begin{equation}}
\def\qe{\end{equation}}
\def\[{\begin{equation}}
\def\]{\end{equation}}

\begin{document}

\begin{frontmatter}

\title{Source waves in the uniform discrete plane with linear and point defects}

\date{\today}

\author
{Anton A. Kutsenko}

\address{Jacobs University, 28759 Bremen, Germany; email: akucenko@gmail.com}

\begin{abstract}
A closed-form expression for the amplitudes of source waves in 2D discrete lattice with local and linear (waveguides) defects is derived. The numerical implementation of this analytic expression is demonstrated by several examples.
\end{abstract}



\end{frontmatter}


{\section{Introduction}\lb{sec1}}

It is well known that the analysis of wave propagation in complex structures with local and linear (waveguides) defects is of great importance in various branches of knowledge: from inverse imaging in industry and medicine to seismic survey in geoscience, see also recent study on metamaterials \cite{GMT2020}. Any analytic results are of particular special interest since they allow us to predict the wave amplitudes at any point with very high precision. However, the analytic solutions of wave equations are known mostly for simple models. For example, 1D structures with point defects can be treated analytically with the help of, e.g., the propagator matrix technique. More complex 2D periodic halfspaces can also be studied by using the advanced multidimensional propagator technique, see, \cite{KS2013}. Further development of this propagator method allows to obtain analytic results for the eigenmodes in the junction of two different half-spaces and in the ``sandwich" structures, see \cite{KKSP1,KKSP2}. The propagator technique is a very powerful tool for the analysis of eigenmodes (guided waves) in discrete and continuous periodic structures with waveguides. However, it is not clear at the moment how to apply this method for the analysis of waves in periodic media with embedded waveguides perturbed by local defects. The presence of point sources can also cause difficulties for the propagator method. Local sources and local defects in the discrete lattice can be treated analytically using the scattering theory, see, e.g., \cite{M1,MC2013}. Recently, some analytic results are obtained for the waves in the discrete uniform media with the embedded two half-lines, see \cite{SM2019}.

\begin{figure}[h]
\center{\includegraphics[width=0.99\linewidth]{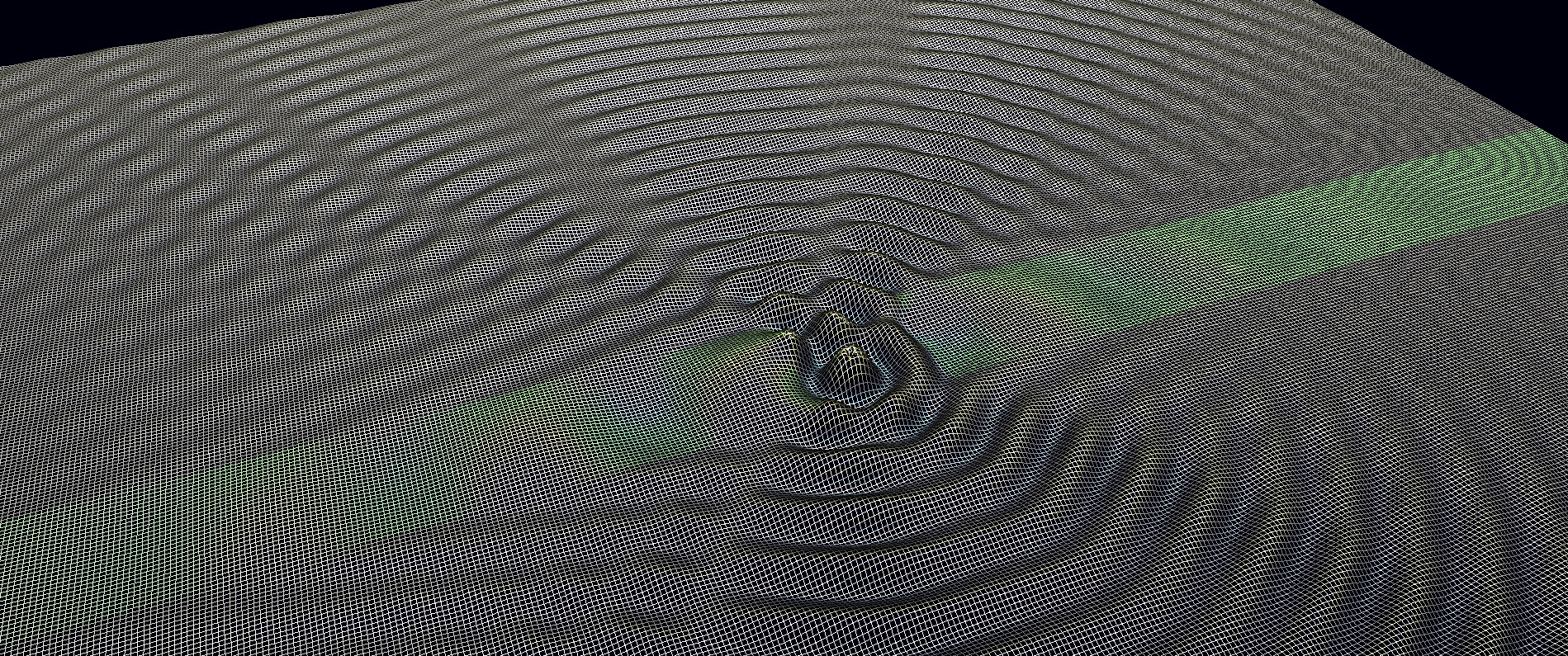}}
\caption{An infinite discrete lattice with two semi-infinite waveguides and one source located between them is illustrated. The wave speed in the lattice is smaller than in the waveguides. Two semi-infinite waveguides can be considered as one infinite waveguide with one local defect. }\lb{fig0}
\end{figure}

In the current paper, we derive a closed-form expression for the waves induced by point sources located in the uniform discrete plane with some local defects and some parallel waveguides. One such configuration is illustrated in Fig. \ref{fig0}. In order to do this we adapt the revised analytic technique developed for self-oscillations in discrete media with defects of various dimensions, see \cite{K1,K3,Kjmaa,Kjmp}. In particular, we extend some of the results from \cite{KIP}, where 2D discrete uniform lattice with local defects but without waveguides is considered. Finally, let us note that the purely infinite plane has some advantages over the finite domains that approximate an infinite one, since there are no spurious waves reflected from artificial boundaries.
  
{\section{Main results}\lb{sec2}}

\subsection{Statement of the problem.} We consider the wave equation on the infinite two-dimensional discrete lattice
\[\lb{001}
 (\D_{\rm discr}U)_{\bf n}=S_{\bf n}\ddot{U}_{\bf n}+\sum_{j=1}^{N_{\rm F}}e^{i\o_j t}\sum_{{\bf m}\in\cN_{{\rm F}j}}F_{{\bf m}j}\d_{{\bf n}{\bf m}},\ \ {\bf n}\in\Z^2,
\]
where $U_{\bf n}\ev U_{\bf n}(t)$ is the amplitude of the wave at the point ${\bf n}$ and time $t$, double dot means the double derivative in time, $\d$ is the Kronecker delta, the discrete Laplace operator is defined by the standard formula
\[\lb{002}
 (\D_{\rm discr}U)_{\bf n}=U_{(n_1+1,n_2)}+U_{(n_1,n_2+1)}+U_{(n_1-1,n_2)}+U_{(n_1,n_2-1)}-4U_{\bf n},\ \ {\bf n}=(n_1,n_2)\in\Z^2,
\]
the finite set $\cN_{{\rm F}j}\ss\Z^2$ is the location of the harmonic forces of the frequency $\o_j\in\C$, their amplitudes $F_{{\bf m}j}\in\C$. Usually, it is enough to assume real amplitudes, but complex amplitudes do not complicate the results. The values $S_{\bf n}=1/C_{\bf n}^2$, where $C_{\bf n}>0$ is the velocity of wave at the point ${\bf n}\in\Z^2$. We assume that the lattice is homogeneous everywhere except for some line and point defects, namely
\[\lb{003}
 S_{\bf n}=\ca 
                 1,& {\bf n}\not\in\cN_{\rm P}\cup\cN_{\rm L},\\
                 1+p_{\bf n},& {\bf n}\in\cN_{\rm P}\sm\cN_{\rm L},\\
                 1+s_{\bf n},& {\bf n}\in\cN_{\rm L}\sm\cN_{\rm P},\\ 
                 1+p_{\bf n}+s_{\bf n},&{\bf n}\in\cN_{\rm P}\cup\cN_{\rm L},
           \ac
\] 
where $\cN_{\rm L}\ss\Z^2$ is a finite set corresponding to the local defects $s_{\bf n}\in\R$, and $\cN_{\rm P}=\Z\ts\cN_{{\rm P}2}$ with some finite $\cN_{{\rm P}2}\ss\Z$ is the location of line defects $p_{(n_1,n_2)}=p_{n_2}\in\R$, $n_2\in\cN_{{\rm P}2}$. Thus, the perturbations $p_{\bf n}$ does not depend on the first coordinate. Note that the locations of line and local defects may have common points, i.e. it is possible that $\cN_{\rm P}\cap\cN_{\rm L}\ne\es$. The locations of the harmonic sources of different frequencies may also intersect each other. They may also intersect with the defect locations.  Our goal is to find analytic formulas for $U_{\bf n}(t)$. Hence, we need to solve infinite number of equations \er{001} explicitly.
 
\subsection{Analytic solution of the wave equation.} We formulate our main result, and postpone its derivation to the next Section. Let $\cN_{\rm R}$ be a finite subset of the lattice $\Z^2$, where we would like to find $U_{\bf n}(t)$ explicitly. This means that we need to find analytic expression for the vector-column 
\[\lb{004}
 {\bf U}(t)=(U_{\bf n}(t))_{{\bf n}\in\cN_{\rm R}}.
\]
For simplicity, let us assume that all the defect perturbations $p_{\bf n}$ and $s_{\bf n}$ are non-zero. This is not a restriction, but this assumption simplifies a little bit some of the next formulas. Firstly, we write the expression for ${\bf U}$ and then define all the ingredients.
\begin{theorem}\lb{T1} Suppose that $S_{\bf n}>0$, ${\bf n}\in\Z_2$; $s_{\bf n}\ne0$, ${\bf n}\in\cN_{\rm S}$; $p_{n_2}\ne0$, $n_2\in\cN_{{\rm P}2}$; and $\Im \o_j\ne0$, $1\le j\le N_{\rm F}$. Then
\[\lb{005}
 {\bf U}(t)=\sum_{j=1}^{N_{\rm F}}e^{i\o_jt}({\bf B}_j(\cN_{\rm R},\cN_{{\rm F}_j})-\o_j^2{\bf B}_j(\cN_{\rm R},\cN_{{\rm L}}){\bf G}_j{\bf B}_j(\cN_{\rm L},\cN_{{\rm F}j})){\bf f}_j,
\]
where the vector-column ${\bf f}_j$ is
\[\lb{006}
 {\bf f}_j=(F_{{\bf n}j})_{{\bf n}\in\cN_{{\rm F}j}}.
\]
The matrices ${\bf B}_j$ are defined by
\[\lb{007}
 {\bf B}_j(\cN,\cM)=\frac1{2\pi}\int_{-\pi}^{\pi}{\bf E}(\cN)({\bf C}_j(\cN,\cM)-\o_j^2{\bf C}_j(\cN,\cN_{{\rm P}0}){\bf F}_j{\bf C}_j(\cN_{{\rm P}0},\cM)){\bf E}^{*}(\cM)dk_1,
\]
where $^*$ denotes the Hermite conjugation, $\cN_{{\rm P}0}=\{0\}\ts\cN_{{\rm P}2}\ss\Z^2$, and $\cN,\cM\ss\Z^2$ are arbitrary finite sets. In turn, the matrices ${\bf G}_j$, ${\bf F}_j$ are given by
\[\lb{008}
 {\bf G}_j=({\bf S}^{-1}+\o_j^2{\bf B}_j(\cN_{\rm L},\cN_{\rm L}))^{-1},\ \ \ {\bf F}_j=({\bf P}^{-1}+\o_j^2{\bf C}_j(\cN_{{\rm P}0},\cN_{{\rm P}0}))^{-1},
\] 
where the diagonal matrices ${\bf S}$ and ${\bf P}$ are defined as
\[\lb{009}
 {\bf S}=\diag(s_{\bf n})_{{\bf n}\in\cN_{\rm L}},\ \ \ {\bf P}=\diag(p_{\bf n})_{{\bf n}\in\cN_{{\rm P}0}}.
\]
Other matrices included in \er{007} and \er{008} are given by
\[\lb{010}
 {\bf E}(\cN)=\diag(e^{-in_1k_1})_{(n_1,n_2)\in\cN},\ \ \ {\bf C}_j(\cN,\cM)=\lt(\frac{z_j^{|n_2-m_2|}}{z_j-z_j^{-1}}\rt)_{(n_1,n_2)\in\cN,(m_1,m_2)\in\cM},
\] 
where $z_j$ is the minimal by the norm value among $\{z_{j-},z_{j+}\}$ given by
\[\lb{011}
 z_{j\pm}=\frac{4-2\cos k_1-\o_j^2\pm\sqrt{(4-2\cos k_1-\o_j^2)^2-4}}2.
\]
\end{theorem}
If $\o_j\not\in\R$ then one of $z_{j\pm}$ has the norm strictly greater than $1$, and the second one lies inside the unit ball, since $z_{j+}z_{j-}=1$. The value $z_j$ in \er{010} coincides with that one lying inside the unit ball. The condition $\o_j\not\in\R$ (more precisely, $\o_j\not\in[-2\sqrt{2},2\sqrt{2}]$, see \cite{K1}) is also necessary for the existence $A^{-1}$. To satisfy $\o_j\not\in\R$, we may assume that $\Im\o_j$ is a very small positive number. This means that the corresponding harmonic source decays very slowly in time. This assumption looks physically reasonable. Moreover, \er{103}, which is equivalent in some sense to \er{001}, see the beginning of Section \ref{sec2}, can be generally written in the form
$$
 (\D_{\rm discr}+\o^2S)V=F\ \ \Rightarrow\ \ V=(\D_{\rm discr}+\o^2S)^{-1}F.
$$ 
The invertibility of $\D_{\rm discr}+\o^2S$ or $S^{-\frac12}\D_{\rm discr}S^{-\frac12}+\o^2$ (recall that $S>0$) guarantees the existence and uniqueness of the solution $V\in\ell^2(\Z^2)$. For the invertibility, it is enough to assume that $\o^2\not\in[0,+\iy)$, since $S^{-\frac12}\D_{\rm discr}S^{-\frac12}$ is negative definite operator with a negative real spectrum. Recall that the Laplace operator is negative definite, and $S^{-\frac12}$ is self-adjoint when $S_{\bf n}>0$, ${\bf n}\in\Z^2$. Thus, the choice of non-real $\o_j$ guarantees also the invertibility of matrices \er{008} and stability of the analytic solution \er{005}.

\subsection{Numerical implementation and examples.} Formula \er{005} is ready for implementation. Most of the operations in \er{005}-\er{011} can be realized with the help of highly optimized linear algebra numerical packages, such as, e.g., Intel MKL, which is a LAPACK implementation optimized for the Intel CPUs and GPUs. We use a LAPACK version of Dew Research Products (MtxVec) based on Intel MKL that provides very efficient numerical libraries for Embarcadero Delphi Community Edition - Object Pascal IDE. Hence, in our implementation, we use the standard BLAS and LAPACK routines only. 

Let us make one remark. The integration \er{007} over $[-\pi,\pi]$ can be reduced to the integral over $[0,\pi]$, since ${\bf C}_j$ and, hence, all  the matrices that expressed through ${\bf C}_j$, are even in $k_1$, see \er{011} and \er{010}. Thus, we can write \er{007} as
\[\lb{012}
 {\bf B}_j(\cN,\cM)=\frac1{\pi}\int_0^{\pi}({\bf C}_j(\cN,\cM)-\o_j^2{\bf C}_j(\cN,\cN_{{\rm P}0}){\bf F}_j{\bf C}_j(\cN_{{\rm P}0},\cM)).\wt {\bf E}(\cN,\cM)dk_1,
\] 
where the symbol $.$ means here the component-by-component product of two matrices, and $\wt{\bf E}$ is given by
\[\lb{013}
 \wt {\bf E}(\cN,\cM)=(\cos(n_1-m_1)k_1)_{(n_1,n_2)\in\cN,(m_1,m_2)\in\cM}.
\]
In the following examples, we use \er{012}-\er{013} instead of \er{007}. 
The reason is that the computation of the integral is a bottleneck in our realization, since, roughly speaking, the corresponding integral thread is outermost by nesting level among optimized matrix operations threads. On the other hand, it should be noted that for other realizations \er{007} may be better than  \er{012}-\er{013}, since instead of taking the complete inverse matrix ${\bf F}_j$ in \er{012} it is enough to solve linear equations of the form $({\bf P}^{-1}+\o_j^2{\bf C}_j(\cN_{{\rm P}0},\cN_{{\rm P}0})){\bf X}=...$ for ${\bf X}$, see \er{005}, \er{007}, and \er{008}.

The integral in \er{012} is computed by taking the standard (uniform) Riemann sum with the number of steps $N_{\rm int}=2000$. For all the examples, we choose the domain $\cN_{\rm R}=[-240,239]^2\ss\Z^2$ with different configurations of local and line defects, and different locations of sources. We draw the real part of the amplitudes ${\bf U}(t)$ for $t\approx0$. Sometimes $t=0.1$ or other value depending on our aesthetic preferences. Of course, when we compare some two figures, we chose the same time for both. Anyway, all the videos for arbitrary time intervals are available by reasonable request to the author of the current paper. 

\subsubsection{Example 0.} We start with the example already considered briefly above, see Fig. \ref{fig0}. The domain for the line defect (waveguide) is determined by $\cN_{{\rm P}2}=[0,30]$, the perturbation of the velocities in this line defect is constant $p_{n_2}=-0.9$, $n_2\in\cN_{{\rm P}2}$. Thus, the wave speed inside the waveguide is $1/\sqrt{1-0.9}\approx3.16$ that is $3$ times faster than the wave speed inside the lattice and inside the local defect defined by $\cN_{\rm S}=[0,30]^2$ and the constant perturbation $s_{\bf n}=0$, ${\bf n}\in\cN_{\rm S}$. There is one harmonic source of the frequency $\o=0.5+0.0005i$ and with the amplitude $f=1$ located at the point $(10,10)\in\Z^2$. 

\subsubsection{Example 1.} We take two sources of the frequency $\o=1+0.0005i$ and the amplitudes
$F_{(0,0)}=1$ and $F_{(80,80)}=-0.5$. The profile of the line defect (channel) is non-constant
\[\lb{014}
 p_{n_2}=-0.9+1.8\sin\frac{\pi (n_2-9.999)}{61},\ \ \ n_2\in\cN_{{\rm P}2}=[10,70].
\]
We compute the wavefield for this configuration and for the same one but with one small defect determined by $\cN_{\rm S}=[-20,-10]\ts[0,10]$ and constant perturbations $s_{\bf n}=-0.9$, ${\bf n}\in\cN_{\rm S}$, see Fig. \ref{fig1}. For better color rendering, we interpolate the grid with triangles in this and next figures, compare with Fig. \ref{fig0}. Readers may zoom figures in on the online version. Even such a small square defect noticeably affects the waves propagating near the corresponding channel bridge closest to the defect.
\begin{figure}[h]
    \centering
    \begin{subfigure}[b]{0.49\textwidth}
        \includegraphics[width=\textwidth]{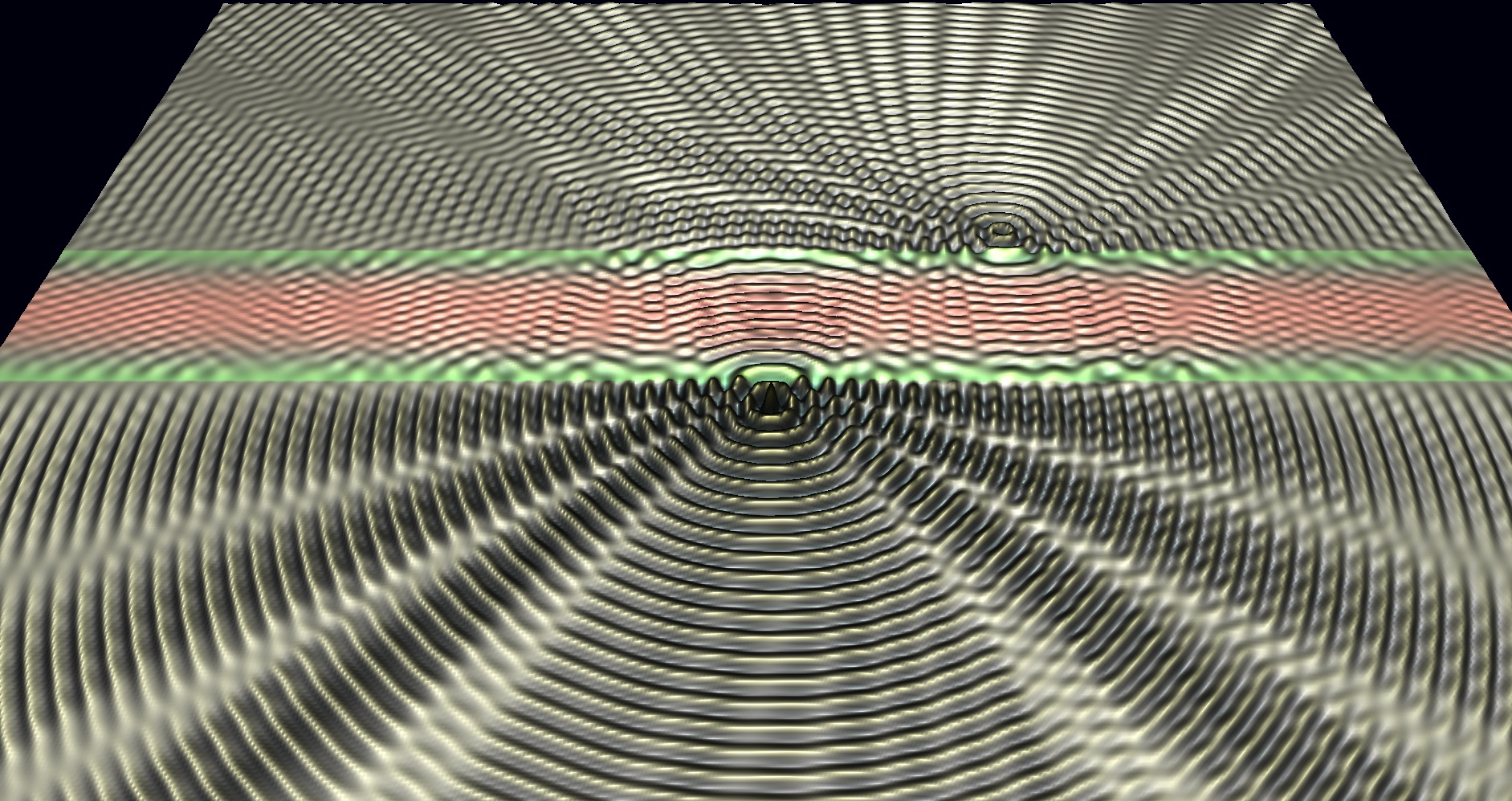}
        \caption{without defect}
    \end{subfigure}
    \begin{subfigure}[b]{0.49\textwidth}
        \includegraphics[width=\textwidth]{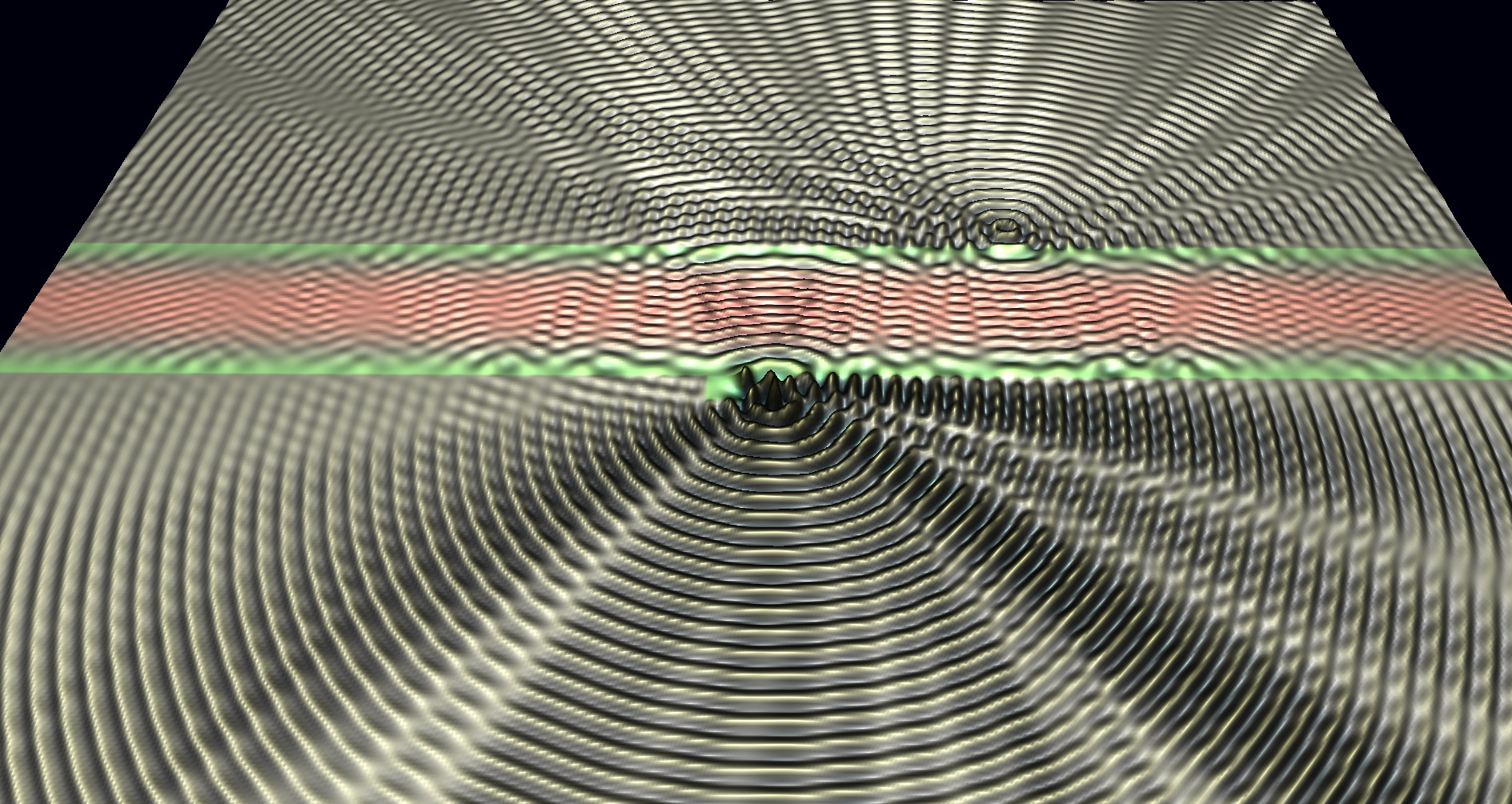}
        \caption{with defect}
    \end{subfigure}
    \caption{Channel (waveguide) is embedded into the homogeneous lattice. Waves come from two sources located near the bridges of the channel. Near one of the sources, we put a small defect, see (b). This defect affects the waves propagating along with the corresponding coast in comparison with the same configuration without the defect.}\lb{fig1}
\end{figure}

\subsubsection{Example 2.} Let us consider two channels $\cN_{{\rm P}2}=[-30,-1]\cup[30,60]$ with the constant perturbation $p_{n_2}=-0.9$, $n_2\in\cN_{{\rm P}2}$. The source of the frequency $\o=1+0.0005i$ is located between the channels $F_{(10,10)}=1$, but at different distances to them. The reflection from the channels increases the amplitudes at the points located between them. The small connection $\cN_{\rm S}=[-100,-80]\ts[0,29]$ between the channels with the same perturbation as in the channels $s_{\bf n}=-0.9$, ${\bf n}\in\cN_{\rm S}$ change wave propagation much stronger than in the previous example, see Fig. \ref{fig2}.
\begin{figure}[h]
    \centering
    \begin{subfigure}[b]{0.49\textwidth}
        \includegraphics[width=\textwidth]{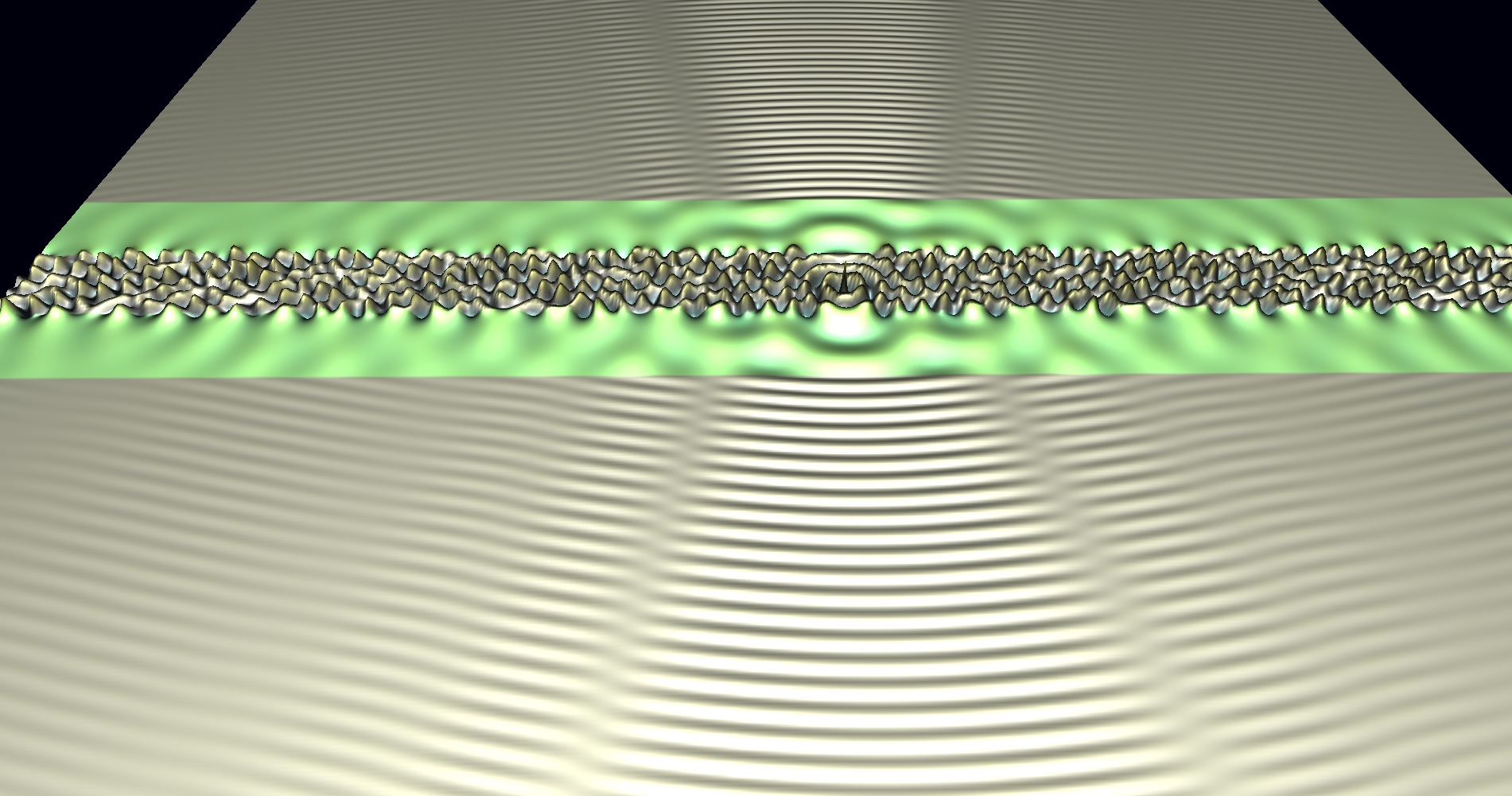}
        \caption{without connection}
    \end{subfigure}
    \begin{subfigure}[b]{0.49\textwidth}
        \includegraphics[width=\textwidth]{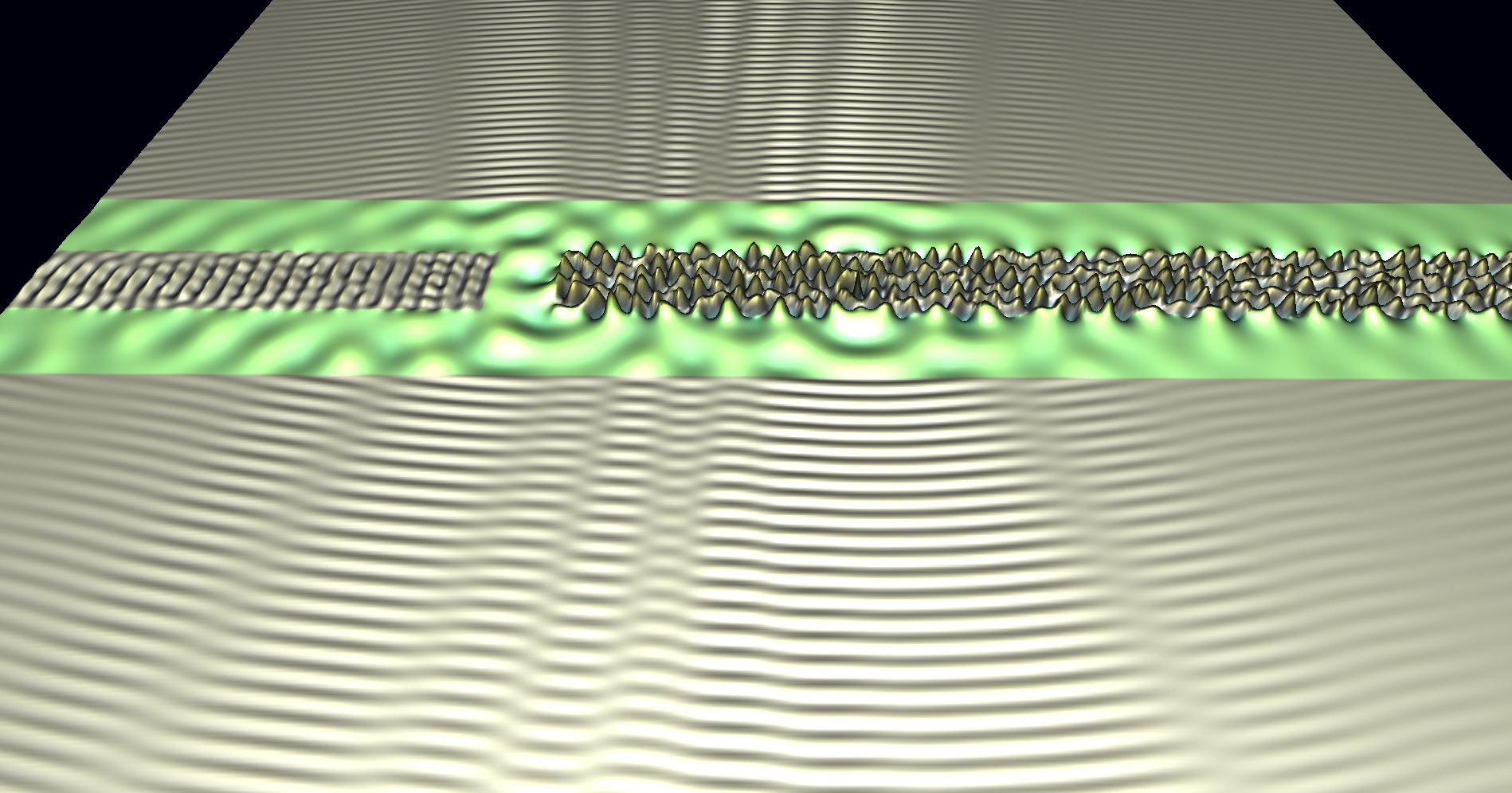}
        \caption{with connection}
    \end{subfigure}
    \caption{Two similar channels are embedded into the homogeneous lattice. The harmonic source is located between them. The connection between the channels changes wave propagation significantly.}\lb{fig2}
\end{figure}

\subsubsection{Example 3.} For some configurations, the effect of a connection between channels is less visible. Let us consider two channels $\cN_{{\rm P}2}=[-30,-1]\cup[30,60]$ with different perturbation $p_{n_2}=-0.9$, $n_2\in[-30,-1]$ and $p_{n_2}=0.9$, $n_2\in[30,60]$. Again, the source of the frequency $\o=1+0.0005i$ is located between the channels $F_{(10,10)}=1$. The small connection $\cN_{\rm S}=[-100,-80]\ts[0,29]$ between the channels with the same perturbation as in the second channel $s_{\bf n}=0.9$, ${\bf n}\in\cN_{\rm S}$ change wave propagation slightly, see Fig. \ref{fig3}.
\begin{figure}[h]
    \centering
    \begin{subfigure}[b]{0.49\textwidth}
        \includegraphics[width=\textwidth]{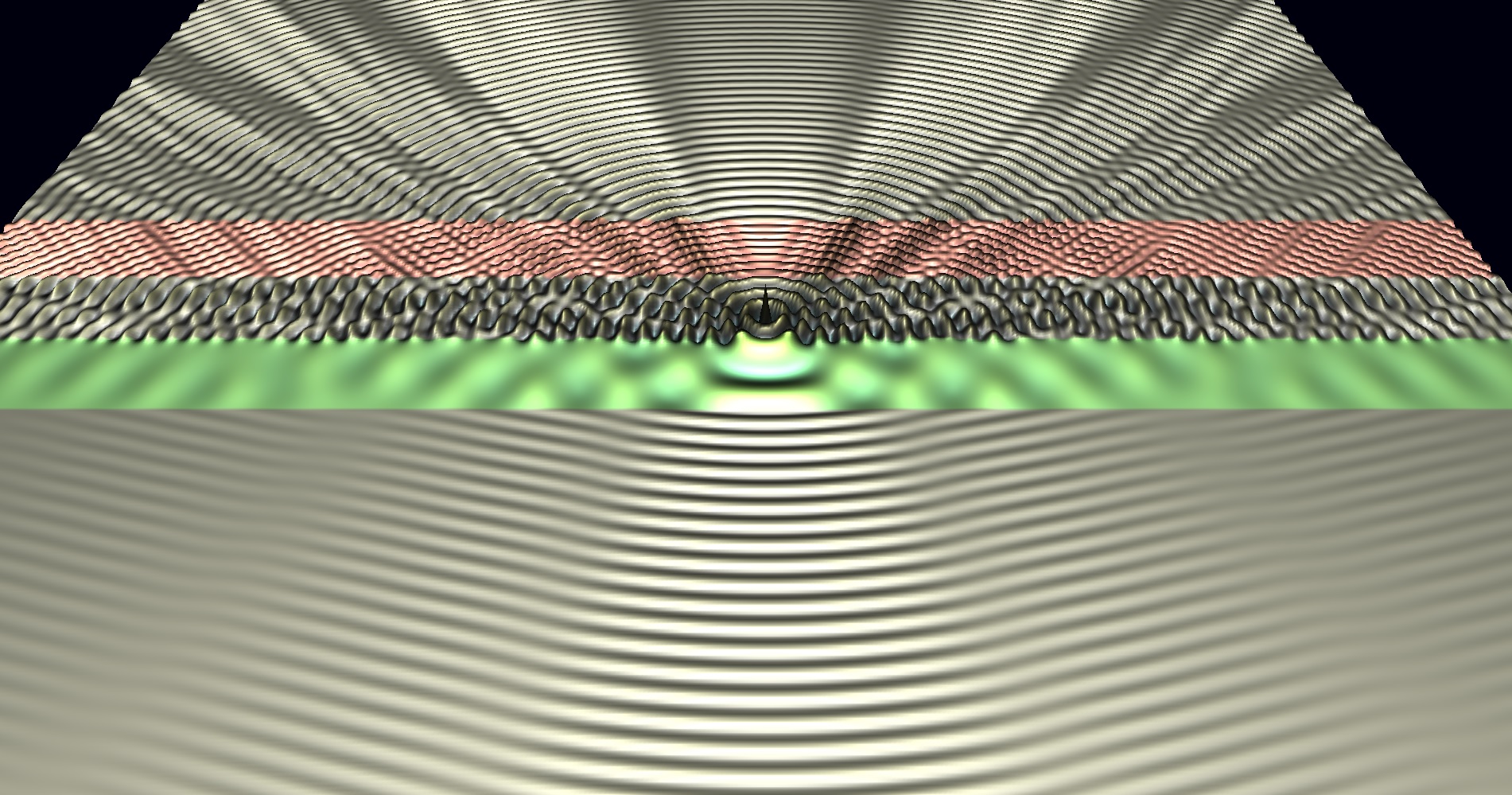}
        \caption{without connection}
    \end{subfigure}
    \begin{subfigure}[b]{0.49\textwidth}
        \includegraphics[width=\textwidth]{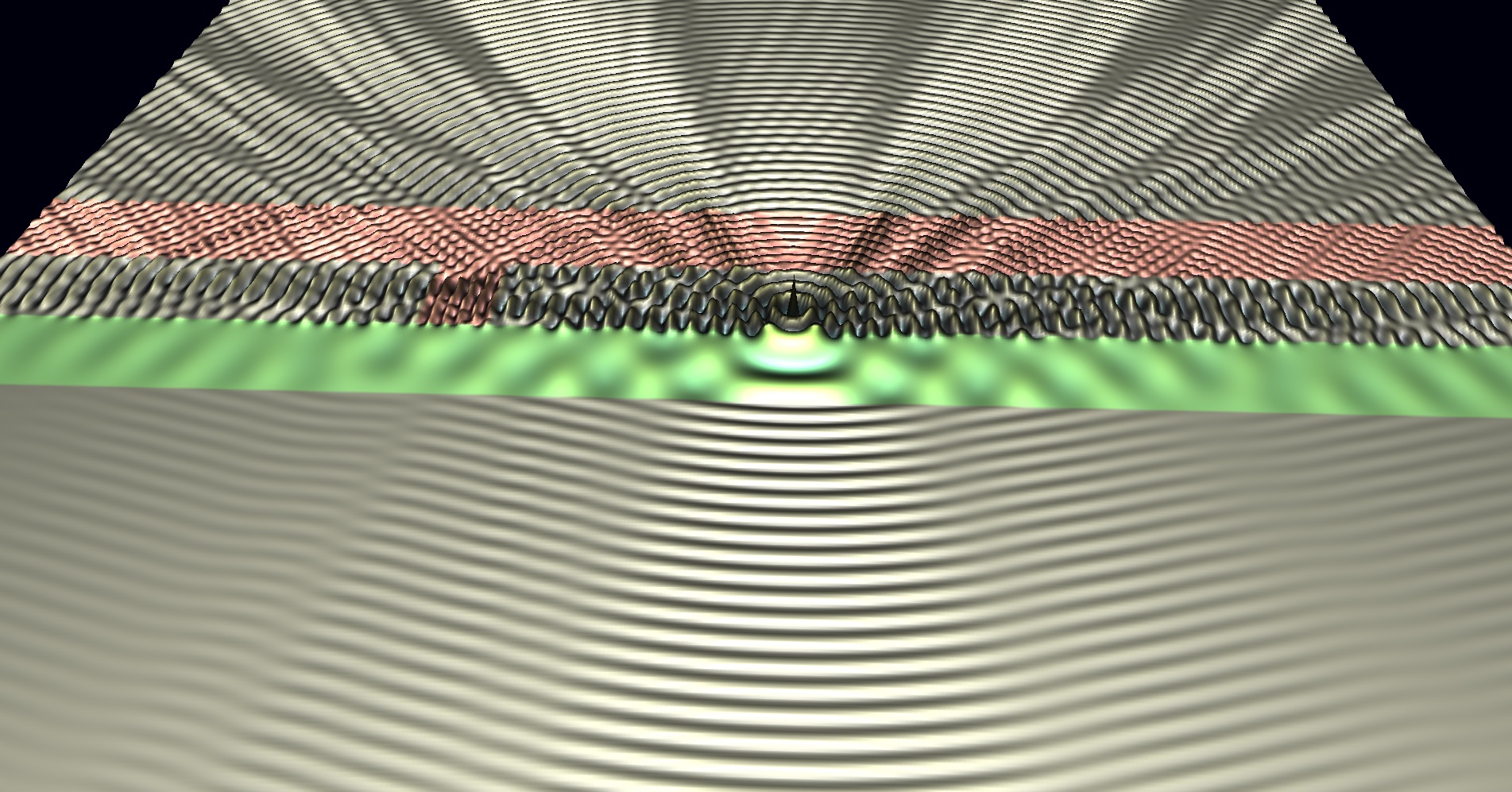}
        \caption{with connection}
    \end{subfigure}
    \caption{Two different channels are embedded into the homogeneous lattice. The harmonic source is located between them. The connection between the channels changes wave propagation slightly.}\lb{fig3}
\end{figure}

\subsubsection{Example 4.} In this example we take the channel with the profile
\[\lb{015}
 p_{n_2}=0.9-1.8\sin\frac{\pi (n_2-9.999)}{61},\ \ \ n_2\in\cN_{{\rm P}2}=[10,70],
\]
similar to \er{014}, but with the opposite sign. The locations of sources $F_{(0,0)}=1$ and $F_{(80,80)}=-0.5$ is the same as in Example 1. Here, we consider two frequencies $\o=1+0.005i$ and $\o=0.5+0.0005i$ for both sources. The local defect area $\cN_{\rm S}=[35,45]\ts[10,70]$ is contained in the channel, the corresponding perturbations make the defect area similar to the channel properties from Example 1, i.e.
\[\lb{016}
 s_{{\bf n}}=2(-0.9+1.8\sin\frac{\pi (n_2-9.999)}{61}),\ \ \ {\bf n}=(n_1,n_2)\in\cN_{{\rm S}}.
\]
The two wavefields related to the configuration without and with the local defect are illustrated in Fig. \ref{fig4}. 
\begin{figure}[h]
    \centering
    \begin{subfigure}[b]{0.49\textwidth}
        \includegraphics[width=\textwidth]{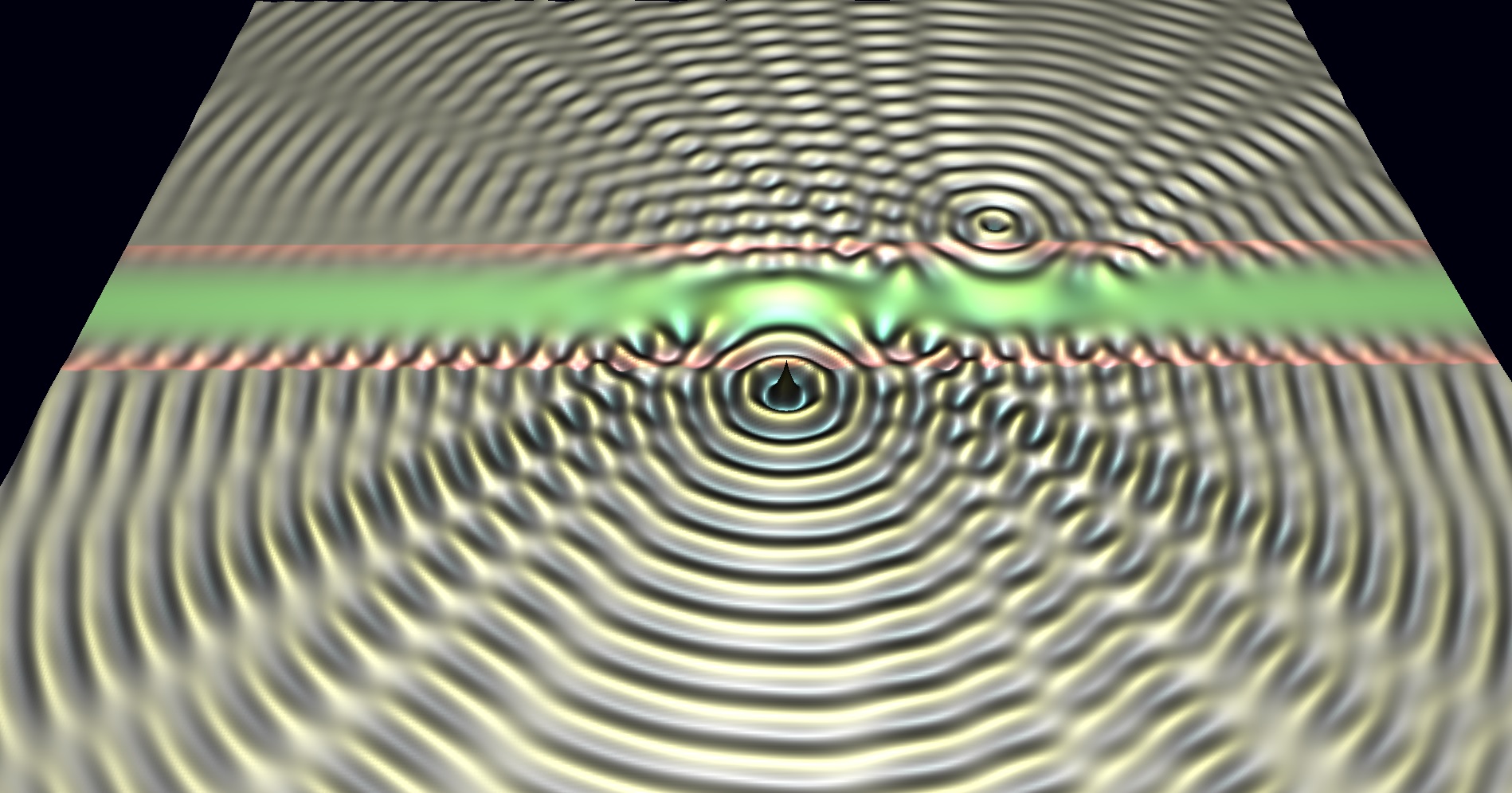}
        \caption{without local defect}
    \end{subfigure}
    \begin{subfigure}[b]{0.49\textwidth}
        \includegraphics[width=\textwidth]{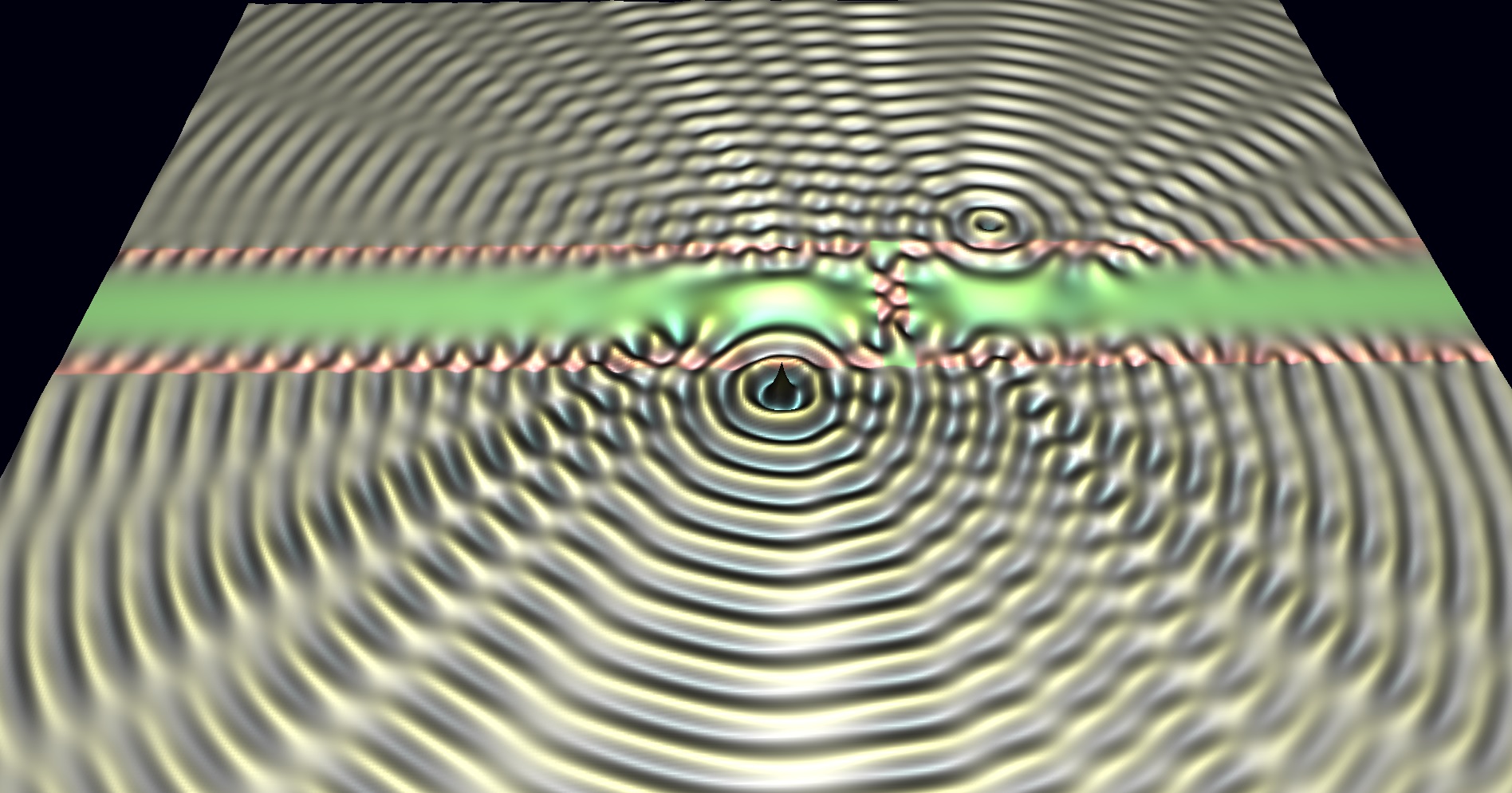}
        \caption{with local defect}
    \end{subfigure}
    \begin{subfigure}[b]{0.49\textwidth}
        \includegraphics[width=\textwidth]{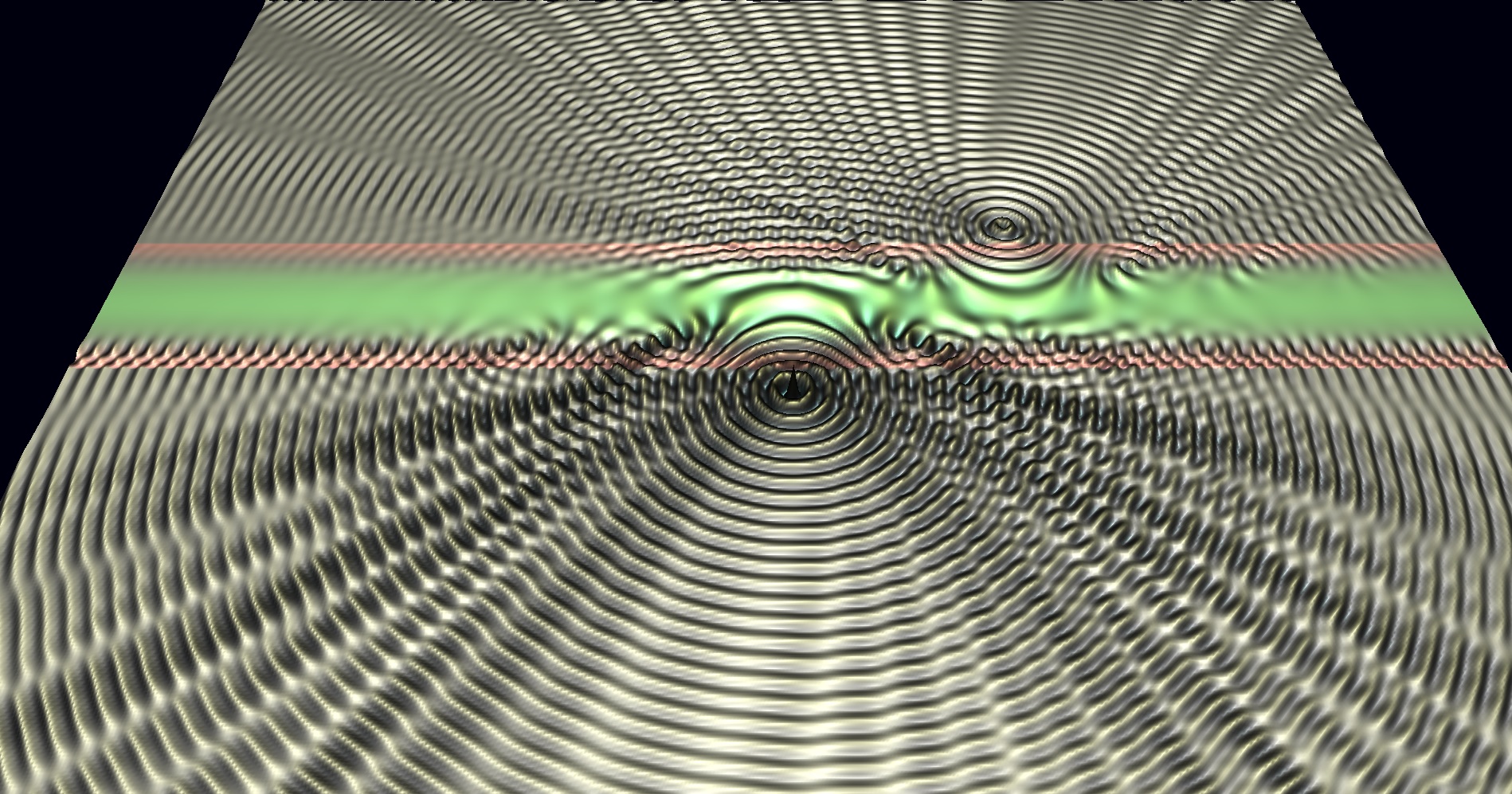}
        \caption{without local defect}
    \end{subfigure}
    \begin{subfigure}[b]{0.49\textwidth}
        \includegraphics[width=\textwidth]{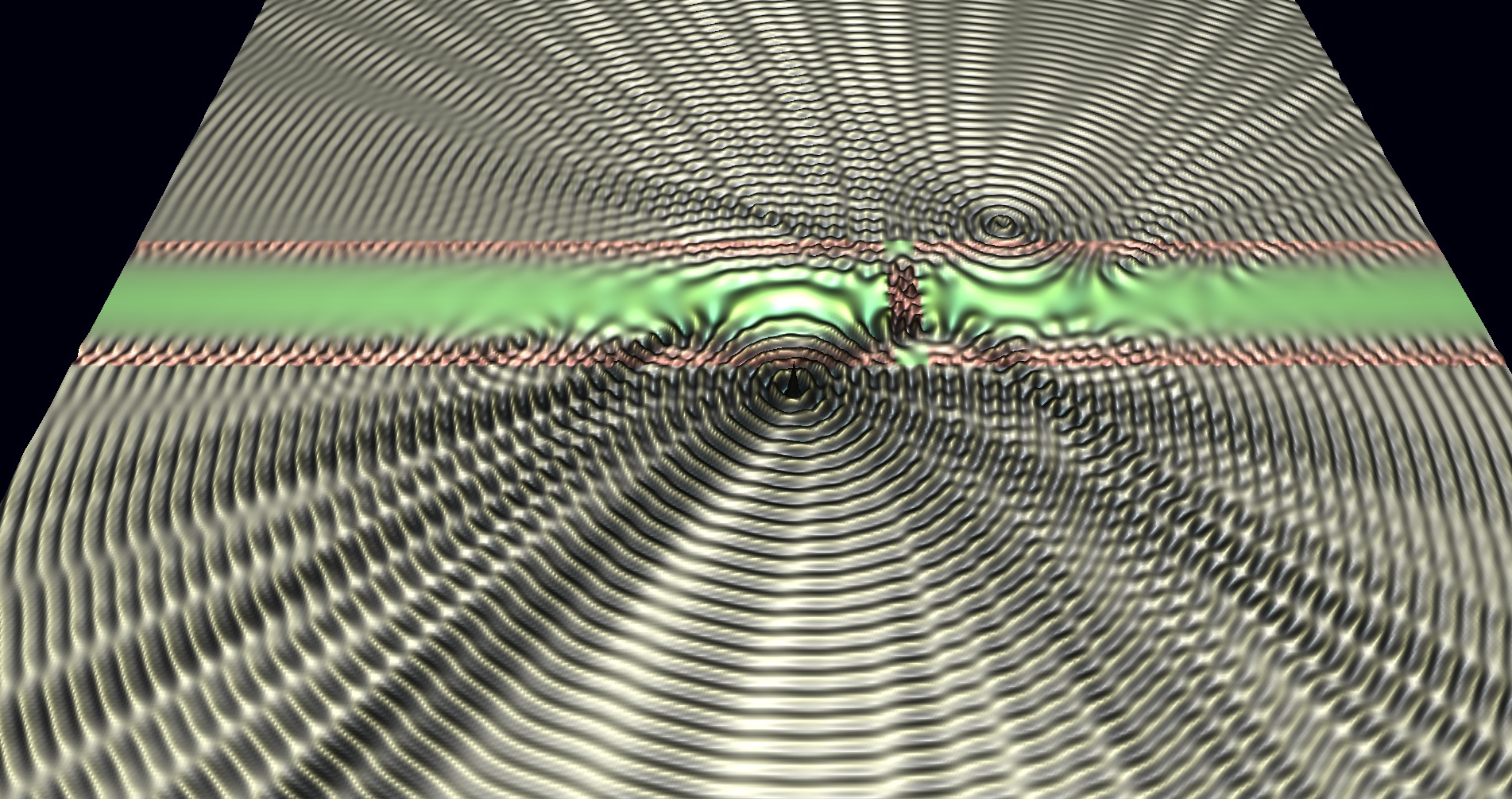}
        \caption{with local defect}
    \end{subfigure}
    \caption{One channels with complex profile \er{015} is embedded into the homogeneous lattice. The defect area ``inverts" the velocities, see \er{16}. Two frequencies are considered, $\o=1+i0.0005$ in (a) and (b), and $\o=0.5+i0.0005$ in (c) and (d).}\lb{fig4}
\end{figure}

\subsubsection{Example 5.} Finally, let us increase the distance between waveguides from Example 2, and change the defect area as well. We consider a rectangular and triangular local defects. The corresponding wavefields are illustrated in Fig. \ref{fig5}. We use a different image format than in the previous examples. It is seen how the waves reflect from the waveguides and how the local defect change wave propagation in general.
\begin{figure}[h]
    \centering
    \begin{subfigure}[b]{0.49\textwidth}
        \includegraphics[width=\textwidth]{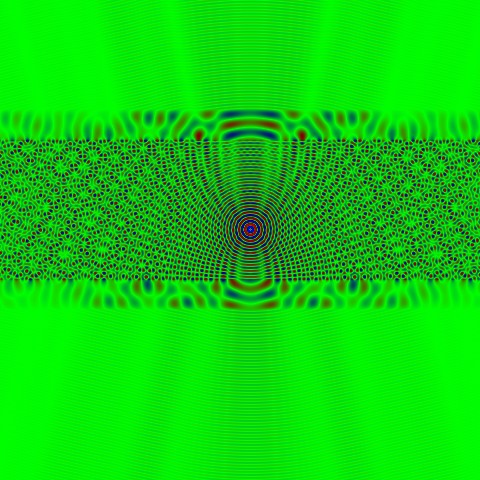}
        \caption{without local defect}
    \end{subfigure}
    \begin{subfigure}[b]{0.49\textwidth}
        \includegraphics[width=\textwidth]{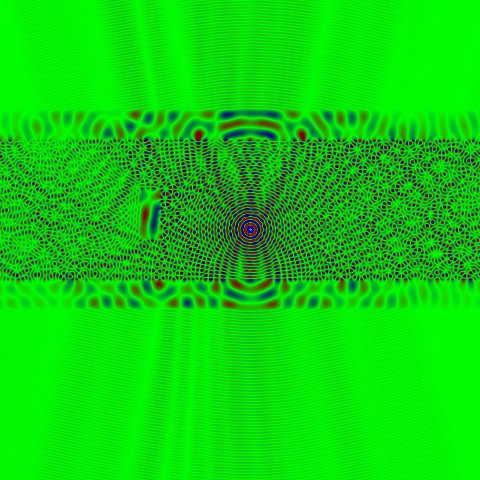}
        \caption{with rectangular}
    \end{subfigure}
    \begin{subfigure}[b]{0.49\textwidth}
        \includegraphics[width=\textwidth]{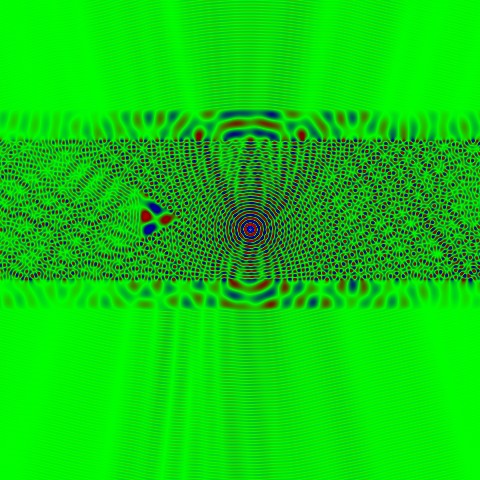}
        \caption{with triangle}
    \end{subfigure}
    \begin{subfigure}[b]{0.49\textwidth}
        \includegraphics[width=\textwidth]{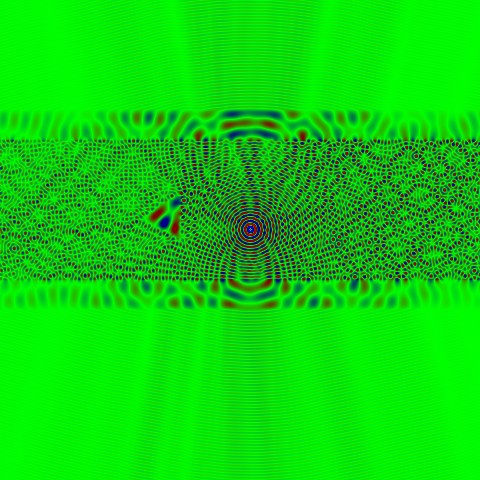}
        \caption{with triangle}
    \end{subfigure}
    \caption{The similar wavefields as in Fig. \ref{fig2}, but with the increased distance between the waveguides, the defect area is also changed.}\lb{fig5}
\end{figure}

{\section{Proof of Theorem \ref{T1}}\lb{sec3}}
Wave equation \er{001} is linear. Hence, due to the superposition principle, the general solution of \er{001} is
\[\lb{100}
 U_{\bf n}=\sum_{j=1}^{N_{\rm F}} U_{{\bf n}j},
\] 
where each $U_{{\bf n}j}$ satisfies
\[\lb{101}
 (\D_{\rm discr}U)_{{\bf n}j}=S_{\bf n}\ddot{U}_{{\bf n}j}+e^{i\o_j t}\sum_{{\bf m}\in\cN_{{\rm F}j}}F_{{\bf m}j}\d_{{\bf n}{\bf m}},\ \ {\bf n}\in\Z^2. 
\]
Hence, without lost of generality, we may fix $1\le j\le N_{\rm F}$ and consider the following wave equation
\[\lb{102}
 (\D_{\rm discr}U)_{{\bf n}}=S_{\bf n}\ddot{U}_{{\bf n}}+e^{i\o t}\sum_{{\bf m}\in\cN_{{\rm F}}}F_{{\bf m}}\d_{{\bf n}{\bf m}},\ \ {\bf n}\in\Z^2, 
\]
where the subscript $j$ is omitted. Now, the standard setting $U_{\bf n}(t)=e^{i\o t}V_{\bf n}$ with $V_{\bf n}\in\C$ rewrites \er{102} as
\[\lb{103}
 (\D_{\rm discr}V)_{{\bf n}}=-\o^2S_{\bf n}{V}_{{\bf n}}+\sum_{{\bf m}\in\cN_{{\rm F}}}F_{{\bf m}}\d_{{\bf n}{\bf m}},\ \ {\bf n}\in\Z^2,
\]
that is an infinite set of linear equations free of time. The next step is the Fourier transform which allows us to rewrite infinite linear system as a finite integral equation. Let us introduce
\[\lb{104}
 v({\bf k})=\sum_{{\bf n}\in\Z^2}V_{\bf n}e^{i{\bf n}\cdot{\bf k}},\ \ {\bf k}=(k_1,k_2)\in[-\pi,\pi]^2.
\] 
The inverse Fourier transform looks like
\[\lb{105}
 V_{\bf n}=\frac1{(2\pi)^2}\int_{-\pi}^{\pi}\int_{-\pi}^{\pi}e^{-i{\bf n}\cdot{\bf k}}v(k_1,k_2)dk_1dk_2=\langle e^{-i{\bf n}\cdot{\bf k}}v\rangle_{21},
\]
where we denote
\[\lb{106}
 \langle...\rangle_1:=\frac1{2\pi}\int_{-\pi}^{\pi}...dk_1,\ \ \langle...\rangle_2:=\frac1{2\pi}\int_{-\pi}^{\pi}...dk_2,\ \ \langle...\rangle_{12}=\langle...\rangle_{21}:=\langle\langle...\rangle_{2}\rangle_1=\langle\langle...\rangle_{1}\rangle_2.
\]
Introducing the vector columns
\[\lb{107}
 {\bf c}=(e^{-in_2k_2})_{n_2\in\cN_{{\rm P}2}},\ \ {\bf a}=(e^{-i{\bf n}\cdot{\bf k}})_{{\bf n}\in\cN_{\rm L}},\ \ {\bf b}=(e^{-i{\bf n}\cdot{\bf k}})_{{\bf n}\in\cN_{{\rm F}}},\ \ {\bf f}=(F_{{\bf n}})_{{\bf n}\in\cN_{{\rm F}}},
\]
and using \er{003}, \er{009} along with \er{104}, \er{105}, we write \er{103} as
\[\lb{eqF}
 Av=-\o^2{\bf a}^*{\bf S}\langle{\bf a}v\rangle_{21}-\o^2{\bf c}^*{\bf P}\langle{\bf c}v\rangle_{2}+{\bf b}^*{\bf f},
\]
where
\[\lb{108}
 A=e^{ik_1}+e^{ik_2}+e^{-ik_1}+e^{-ik_2}-4+\o^2=\o^2-4+2\cos k_1+2\cos k_2
\]
is the form of $\D_{\rm discr}+\o^2$ in the Fourier space. The goal is to find the solution $v$ of \er{eqF} explicitly. The first step in this way is to express $\langle{\bf c}v\rangle_{2}$. Using \er{eqF}, we obtain
\[\lb{cv1}
 {\bf c}v=-\o^2\frac{{\bf c}{\bf a}^*}{A}{\bf S}\langle{\bf a}v\rangle_{21}-\o^2\frac{{\bf c}{\bf c}^*}{A}{\bf P}\langle{\bf c}v\rangle_{2}+\frac{{\bf c}{\bf b}^*}{A}{\bf f},
\]
which after integration $\langle...\rangle_2$ gives
\[\lb{cv2}
 \langle{\bf c}v\rangle_2=-\o^2\lt\langle\frac{{\bf c}{\bf a}^*}{A}\rt\rangle_2{\bf S}\langle{\bf a}v\rangle_{21}-\o^2\lt\langle\frac{{\bf c}{\bf c}^*}{A}\rt\rangle_2{\bf P}\langle{\bf c}v\rangle_{2}+\lt\langle\frac{{\bf c}{\bf b}^*}{A}\rt\rangle_2{\bf f},
\]
that, in turn, leads to
\[\lb{cv3}
 \lt({\bf I}+\o^2\lt\langle\frac{{\bf c}{\bf c}^*}{A}\rt\rangle_2{\bf P}\rt)\langle{\bf c}v\rangle_2=-\o^2\lt\langle\frac{{\bf c}{\bf a}^*}{A}\rt\rangle_2{\bf S}\langle{\bf a}v\rangle_{21}+\lt\langle\frac{{\bf c}{\bf b}^*}{A}\rt\rangle_2{\bf f},
\]
and, hence,
\[\lb{cv3a}
 \langle{\bf c}v\rangle_2=-\o^2{\bf D}\lt\langle\frac{{\bf c}{\bf a}^*}{A}\rt\rangle_2{\bf S}\langle{\bf a}v\rangle_{21}+{\bf D}\lt\langle\frac{{\bf c}{\bf b}^*}{A}\rt\rangle_2{\bf f},
\]
where
\[\lb{D}
  {\bf D}:=\lt({\bf I}+\o^2\lt\langle\frac{{\bf c}{\bf c}^*}{A}\rt\rangle_2{\bf P}\rt)^{-1}.
\]
Substituting $\langle{\bf c}v\rangle_{2}$ from \er{cv3} back into \er{eqF}, we obtain
\begin{multline}\lb{eqF1}
 Av=-\o^2{\bf a}^*{\bf S}\langle{\bf a}v\rangle_{21}-\o^2{\bf c}^*{\bf P}\lt(-\o^2{\bf D}\lt\langle\frac{{\bf c}{\bf a}^*}{A}\rt\rangle_2{\bf S}\langle{\bf a}v\rangle_{21}+{\bf D}\lt\langle\frac{{\bf c}{\bf b}^*}{A}\rt\rangle_2{\bf f}\rt)+{\bf b}^*{\bf f}=\\
 \lt(-\o^2{\bf a}^*+\o^4{\bf c}^*{\bf P}{\bf D}\lt\langle\frac{{\bf c}{\bf a}^*}{A}\rt\rangle_2\rt){\bf S}\langle{\bf a}v\rangle_{21}+\lt(-\o^2{\bf c}^*{\bf P}{\bf D}\lt\langle\frac{{\bf c}{\bf b}^*}{A}\rt\rangle_2+{\bf b}^*\rt){\bf f}.
\end{multline}
Now, we express $\langle{\bf a}v\rangle_{21}$. Using \er{eqF1}, we obtain
\[\lb{av1}
 {\bf a}v=\lt(-\o^2\frac{{\bf a}{\bf a}^*}{A}+\o^4\frac{{\bf a}{\bf c}^*}{A}{\bf P}{\bf D}\lt\langle\frac{{\bf c}{\bf a}^*}{A}\rt\rangle_2\rt){\bf S}\langle{\bf a}v\rangle_{21}+\lt(-\o^2\frac{{\bf a}{\bf c}^*}{A}{\bf P}{\bf D}\lt\langle\frac{{\bf c}{\bf b}^*}{A}\rt\rangle_2+\frac{{\bf a}{\bf b}^*}{A}\rt){\bf f},
\]
which after integration $\langle...\rangle_2$ gives 
\begin{multline}\lb{av2}
 \langle{\bf a}v\rangle_2=\lt(-\o^2\lt\langle\frac{{\bf a}{\bf a}^*}{A}\rt\rangle_2+\o^4\lt\langle\frac{{\bf a}{\bf c}^*}{A}\rt\rangle_2{\bf P}{\bf D}\lt\langle\frac{{\bf c}{\bf a}^*}{A}\rt\rangle_2\rt){\bf S}\langle{\bf a}v\rangle_{21}+\\
 \lt(-\o^2\lt\langle\frac{{\bf a}{\bf c}^*}{A}\rt\rangle_2{\bf P}{\bf D}\lt\langle\frac{{\bf c}{\bf b}^*}{A}\rt\rangle_2+\lt\langle\frac{{\bf a}{\bf b}^*}{A}\rt\rangle_2\rt){\bf f}.
\end{multline}
Another integration $\langle...\rangle_1$ turns \er{av2} into
\begin{multline}\lb{eqF4a}
 \langle{\bf a}v\rangle_{21}=\lt(-\o^2\lt\langle\frac{{\bf a}{\bf a}^*}{A}\rt\rangle_{21}+\o^4\lt\langle\lt\langle\frac{{\bf a}{\bf c}^*}{A}\rt\rangle_2{\bf P}{\bf D}\lt\langle\frac{{\bf c}{\bf a}^*}{A}\rt\rangle_2\rt\rangle_1\rt){\bf S}\langle{\bf a}v\rangle_{21}+\\
 \lt(-\o^2\lt\langle\lt\langle\frac{{\bf a}{\bf c}^*}{A}\rt\rangle_2{\bf P}{\bf D}\lt\langle\frac{{\bf c}{\bf b}^*}{A}\rt\rangle_2\rt\rangle_1+\lt\langle\frac{{\bf a}{\bf b}^*}{A}\rt\rangle_{21}\rt){\bf f},
\end{multline}
which leads to
\[\lb{av3}
 \langle{\bf a}v\rangle_{21}={\bf H}\lt(-\o^2\lt\langle\lt\langle\frac{{\bf a}{\bf c}^*}{A}\rt\rangle_2{\bf P}{\bf D}\lt\langle\frac{{\bf c}{\bf b}^*}{A}\rt\rangle_2\rt\rangle_1+\lt\langle\frac{{\bf a}{\bf b}^*}{A}\rt\rangle_{21}\rt){\bf f},
\]
where 
\[\lb{H}
 {\bf H}:=\lt({\bf I}+\o^2\lt\langle\frac{{\bf a}{\bf a}^*}{A}\rt\rangle_{21}{\bf S}-\o^4\lt\langle\lt\langle\frac{{\bf a}{\bf c}^*}{A}\rt\rangle_2{\bf P}{\bf D}\lt\langle\frac{{\bf c}{\bf a}^*}{A}\rt\rangle_2\rt\rangle_1{\bf S}\rt)^{-1}.
\]
Substituting \er{av3} back into \er{eqF1}, we finally obtain
\begin{multline}\lb{eqF2}
 v=\lt(-\o^2\frac{{\bf a}^*}{A}+\o^4\frac{{\bf c}^*}{A}{\bf P}{\bf D}\lt\langle\frac{{\bf c}{\bf a}^*}{A}\rt\rangle_2\rt){\bf S}{\bf H}\lt(-\o^2\lt\langle\lt\langle\frac{{\bf a}{\bf c}^*}{A}\rt\rangle_2{\bf P}{\bf D}\lt\langle\frac{{\bf c}{\bf b}^*}{A}\rt\rangle_2\rt\rangle_1+\lt\langle\frac{{\bf a}{\bf b}^*}{A}\rt\rangle_{21}\rt){\bf f}\\
 +\lt(-\o^2\frac{{\bf c}^*}{A}{\bf P}{\bf D}\lt\langle\frac{{\bf c}{\bf b}^*}{A}\rt\rangle_2+\frac{{\bf b}^*}{A}\rt){\bf f},
\end{multline}
or,
\begin{multline}\lb{eqF3}
 v=\o^2\lt(\o^2\frac{{\bf c}^*}{A}{\bf P}{\bf D}\lt\langle\frac{{\bf c}{\bf a}^*}{A}\rt\rangle_2-\frac{{\bf a}^*}{A}\rt){\bf S}{\bf H}\lt(\lt\langle\frac{{\bf a}{\bf b}^*}{A}\rt\rangle_{21}-\o^2\lt\langle\lt\langle\frac{{\bf a}{\bf c}^*}{A}\rt\rangle_2{\bf P}{\bf D}\lt\langle\frac{{\bf c}{\bf b}^*}{A}\rt\rangle_2\rt\rangle_1\rt){\bf f}\\
 +\lt(\frac{{\bf b}^*}{A}-\o^2\frac{{\bf c}^*}{A}{\bf P}{\bf D}\lt\langle\frac{{\bf c}{\bf b}^*}{A}\rt\rangle_2\rt){\bf f},
\end{multline}
that is the explicit form of the solution of the initial wave equation \er{eqF}.

Recall that $\cN_{\rm R}=\{{\bf n}_i\}_{i=1}^R\ss\Z^2$ is the set of points at which we would like to compute the field at time $t$. The field at $\cN_{\rm R}$ is the array of amplitudes multiplied by $e^{i\o t}$, by the notation above \er{103}, it is $e^{i\o t}(V_{\bf n})_{{\bf n}\in\cN_{\rm R}}$. Due to \er{eqF3} and \er{105}, the field at $\cN_{\rm R}$ is
\begin{multline}\lb{eqF4}
 e^{i\o t}\langle {\bf r}v\rangle_{21}=e^{i\o t}\o^2\lt(\o^2\lt\langle\lt\langle\frac{{\bf r}{\bf c}^*}{A}\rt\rangle_2{\bf P}{\bf D}\lt\langle\frac{{\bf c}{\bf a}^*}{A}\rt\rangle_2\rt\rangle_1-\lt\langle\frac{{\bf r}{\bf a}^*}{A}\rt\rangle_{21}\rt){\bf S}{\bf H}\cdot\\
 \lt(\lt\langle\frac{{\bf a}{\bf b}^*}{A}\rt\rangle_{21}-\o^2\lt\langle\lt\langle\frac{{\bf a}{\bf c}^*}{A}\rt\rangle_2{\bf P}{\bf D}\lt\langle\frac{{\bf c}{\bf b}^*}{A}\rt\rangle_2\rt\rangle_1\rt){\bf f}
 +\\
 e^{i\o t}\lt(\lt\langle\frac{{\bf r}{\bf b}^*}{A}\rt\rangle_{21}-\o^2\lt\langle\lt\langle\frac{{\bf r}{\bf c}^*}{A}\rt\rangle_2{\bf P}{\bf D}\lt\langle\frac{{\bf c}{\bf b}^*}{A}\rt\rangle_2\rt\rangle_1\rt){\bf f},\ \ \ where\ \ \ {\bf r}=\ma e^{-i{\bf n}_1\cdot{\bf k}}\\ ...\\ e^{-i{\bf n}_R\cdot{\bf k}}\am.
\end{multline}
For convenience, we denote
\[\lb{RCA}
 {\bf A}({\bf x},{\bf z})=\lt\langle\frac{{\bf x}{\bf z}^*}{A}\rt\rangle_{21}-\o^2\lt\langle\lt\langle\frac{{\bf x}{\bf c}^*}{A}\rt\rangle_2{\bf P}{\bf D}\lt\langle\frac{{\bf c}{\bf z}^*}{A}\rt\rangle_2\rt\rangle_1.
\]
Recall that all the velocities perturbations are non-zero, so that ${\bf P}^{-1}$ and ${\bf S}^{-1}$ are correctly defined. Then, using \er{D}, \er{H} and \er{RCA}, we obtain
\[\lb{G}
 {\bf G}:={\bf S}{\bf H}=\lt({\bf S}^{-1}+\o^2{\bf A}({\bf a},{\bf a})\rt)^{-1}
\]
and
\[\lb{RCA1}
 {\bf A}({\bf x},{\bf z})=\lt\langle\frac{{\bf x}{\bf z}^*}{A}\rt\rangle_{21}-\o^2\lt\langle\lt\langle\frac{{\bf x}{\bf c}^*}{A}\rt\rangle_2{\bf F}\lt\langle\frac{{\bf c}{\bf z}^*}{A}\rt\rangle_2\rt\rangle_1,
\]
where
\[\lb{F}
 {\bf F}:={\bf P}{\bf D}=\lt({\bf P}^{-1}+\o^2\lt\langle\frac{{\bf c}{\bf c}^*}{A}\rt\rangle_2\rt)^{-1}.
\]
The field \er{eqF4} can also be written more compactly
\[\lb{eqF5}
 e^{i\o t}\langle{\bf r}v\rangle_{21}=e^{i\o t}({\bf A}({\bf r},{\bf b})-\o^2{\bf A}({\bf r},{\bf a}){\bf G}{\bf A}({\bf a},{\bf b})){\bf f}.
\]
The computation of ${\bf A}$, see \er{RCA1}, is most expensive part in numerical implementation of \er{eqF5}, since \er{RCA1} contains a double integration. Fortunately, one of the integrals can be computed explicitly. Namely,
\[\lb{201}
 \lt\langle\frac{e^{i nk_2}}{A}\rt\rangle_2=\frac1{2\pi}\int_{-\pi}^{\pi}\frac{e^{i nk_2}}{A}dk_2=\frac1{2\pi}\int_{-\pi}^{\pi}\frac{e^{i nk_2}}{\o^2-4+2\cos k_1+2\cos k_2}dk_2=\frac{z^{|n_2|}}{z-z^{-1}},
\]
where $z$ is defined above \er{011} ($\o_j$ is replaced by $\o$). The derivation of \er{201} is sufficiently simple and can be obtained by different ways, one is given in formulas (28) and (29) of \cite{K3}. Let
\[\lb{202}
 {\bf x}=(e^{-i{\bf n}\cdot{\bf k}})_{{\bf n}\in\cN},\ \ \ {\bf z}=(e^{-i{\bf m}\cdot{\bf k}})_{{\bf m}\in\cM}
\] 
be two vector columns, where $\cN,\cM\ss\Z^2$ are some finite sets. Then, using \er{201} and \er{202} along with definitions \er{010}, we get
\[\lb{203}
 \lt\langle\frac{xz^*}{A}\rt\rangle_2={\bf E}(\cN){\bf C}(\cN,\cM){\bf E}^*(\cM),
\]
where the subscript $j$ is omitted. Applying \er{203} to \er{RCA1}, using definitions \er{007}, \er{106}, and the fact that ${\bf E}(\cN_{{\rm P}0})$ is the identity matrix, see \er{010}, we obtain ${\bf A}({\bf x},{\bf z})={\bf B}(\cN,\cM)$, where, again, the subscript $j$ is omitted. Similarly, ${\bf G}$ and ${\bf F}$ in \er{G} and \er{F} coincide with that ones in \er{008}. Thus, \er{005} follows from \er{eqF5} and \er{100}. 

\section*{Acknowledgements} 
This paper is a contribution to the project M3 of the Collaborative Research Centre TRR 181 ``Energy Transfer in Atmosphere and Ocean'' funded by the Deutsche Forschungsgemeinschaft (DFG, German Research Foundation) - Projektnummer 274762653. 

\bibliography{bibl_perp1}

\end{document}